\pdfoutput=1
\documentclass[aps,prb,twocolumn,floatfix,amsmath,amssymb,superscriptaddress]{revtex4}
\usepackage[final]{graphicx}
\usepackage{bm}
\usepackage{afterpage}
\usepackage{color}
\usepackage{comment}
\usepackage[colorlinks=true,urlcolor=blue,linkcolor=blue,citecolor=blue]{hyperref}

\begin{document}

\title{Universal approach to $p$-wave triplet superconductivity in the Hubbard models}

\author{Wanpeng Han}
\affiliation{School of Physics, Beihang University,
Beijing, 100191, China}

\author{Xingchuan Zhu }

\affiliation{Interdisciplinary Center for Fundamental and Frontier Sciences, Nanjing University of Science and Technology, Jiangyin, Jiangsu 214443, P. R. China}

\author{Shiping Feng}
\affiliation{ Department of Physics,  Beijing Normal University, Beijing, 100875, China}

\author{Huaiming Guo}
\email{hmguo@buaa.edu.cn}
\affiliation{School of Physics, Beihang University,
Beijing, 100191, China}

\begin{abstract}
Spin-triplet superconductivity is actively pursued in condensed matter physics due to the potential applications in topological quantum computations. The related pairing mechanism involving the interaction remains an important research topic. Here we propose a universal approach to obtain $p$-wave triplet superconductivity in the Hubbard models by simply changing the sign of the hopping amplitudes of the spin-down electrons, and apply it to three prototype two-dimensional lattices (honeycomb, square, and triangular). The parent Hamiltonian at half filling has long-range magnetic order, which is ferromagnetic in all three directions for the frustrated triangular lattices, and ferromagnetic (antiferromagnetic) in the $xy$ plane ($z$ direction) for the bipartite honeycomb and square lattices. The magnetic transitions occur at some critical interactions on honeycomb and triangular lattices, which are estimated by finite-size scalings. When the systems are doped, we find the triplet $p$-wave pairing is a dominating superconducting instability. We demonstrate its emergence is closely related to the strong ferromagnetic spin fluctuations induced by the doping. Our results provide an understanding of the microscopical triplet-pairing mechanism, and will be helpful in the search for spin-triplet superconducting materials.
\end{abstract}


\maketitle
\section{Introduction}
Superconductivity is one of the most fundamental phenomena in condensed matter physics, and has attracted great interest from understanding the fundamental physics to the practical applications. While most superconductors have spin-singlet pairings, such as: BCS superconductors\cite{PhysRev.108.1175}, high-temperature cuprate superconductors\cite{RevModPhys.78.17}, et al., the spin-triplet pairing is only reported in a few very rare cases including $\text{UPt}_3$\cite{doi:10.1126/science.1187943}, $\text{UTe}_2$\cite{doi:10.1126/science.aav8645}, and the two distinct superfluid phases of $^3\text{He}$\cite{RevModPhys.47.331,RevModPhys.75.657}.

Spin-triplet superconductivity is of great interest due to their intriguing physical properties. Spin-triplet superconductors naturally exhibit topological superconductivity, in which Majorana bound states may appear inside the vortex cores\cite{PhysRevB.61.10267}. Currently, Majorana fermions in the solid-state setups are actively pursued due to their potential applications in topological quantum computation\cite{Alicea_2012,doi:10.1146/annurev-conmatphys-030212-184337,Stanescu_2013,RevModPhys.87.137}. The $p$-wave superconductors are more commonly  engineered in the mesoscale systems with spin-orbit couplings by superconducting proximity effect. It has been proposed that a two-dimensional $p$-wave superconductor could arise at an interface between an $s$-wave superconductor and a strong topological insulator\cite{PhysRevLett.100.096407}. In one dimension, the $p$-wave superconductivity, described by the celebrated Kitaev chain
\cite{Kitaev_2001}, can be created based on the edge of a two-dimensional topological insulator, nanowires made of a three-dimensional topological insulator\cite{PhysRevB.84.201105}, semiconductor quantum wires with strong spin-orbit coupling\cite{PhysRevLett.105.077001,PhysRevLett.105.177002}, and helical spin chains\cite{PhysRevB.84.195442,PhysRevB.85.144505}. The above theoretically proposed platforms have been the focus of experimental studies, and great progress in fabricating the hybrid-structure devices and detecting the signature of topological superconductivity has been achieved so far\cite{doi:10.1126/science.1259327,doi:10.1126/science.1222360,PhysRevLett.112.217001}.

Many materials platforms discovered recently may be candidates for spin-triplet Cooper pairings. It is shown that magic-angle twisted trilayer graphene exhibits superconductivity up to extremely large in-plane magnetic field, which violates the Pauli limit for spin-singlet superconductivity, thus suggesting a possibility of a spin-triplet order parameter therein\cite{cao2021pauli}. A three-particle mechanism for spin-triplet superconductivity is presented in multiband systems, according to which the recently discovered dilute superconductors such as $\text{ZrNCl}$, $\text{WTe}_2$ are proposed to be spin-triplet\cite{doi:10.1073/pnas.2117735119}. From the unchanged spin susceptibility with the magnetic field, $\text{K}_2\text{Cr}_3\text{As}_3$ single crystal is established as a spin-triplet superconductor\cite{yang2021spin}. Furthermore, proximity-induced spin-triplet superconductivity is reported in the topological Kagome metal $\text{K}_{1-x}\text{V}_3\text{Sb}_5$\cite{wang2020proximity}.

Although significant achievements have been obtained in investigating spin-triplet superconductivity, how the pairing microscopically originates from many-body interactions is still less understood. In this paper, we provide a universal approach to obtain $p$-wave triplet superconductivity in the Hubbard models by simply changing the sign of the hopping amplitudes of the spin-down electrons. We call the resulting Hamiltonian the imbalanced Hubbard model, and study it using determinant quantum Monte Carlo (DQMC) on the honeycomb, square, and triangular lattices. The magnetic transitions at half filling are first investigated by calculating the spin structure factor. While the frustrated triangular lattice has ferromagnetic (FM) orders in all three direction, the bipartite lattice (honeycomb or square) exhibits a FM and an antiferromagnetic (AF) ones in the $xy$-plane and the $z$ direction, respectively. The magnetic order of the honeycomb lattice and the $xy$-plane FM order of the triangular lattice occur at some critical interactions, which are determined by finite-size scaling analyses of the structure factors. We then calculate the effective pairing susceptibility in the doped systems, and find all the three considered lattices have $p$-wave superconducting instabilities. Finally, we discuss the relation between the FM fluctuations and the triplet superconducting pairings.

This paper is organized as follows. Section II introduces the model we will investigate, along with our computational methodology.
Section III presents the results of the magnetic transitions in the imbalanced honeycomb-, square-, and triangle-lattice Hubbard models. Section IV demonstrates the $p$-wave triplet superconductivity in the corresponding doped Hamiltonians. Section V makes the conclusions.

\begin{figure}[htbp]
\centering \includegraphics[width=8cm]{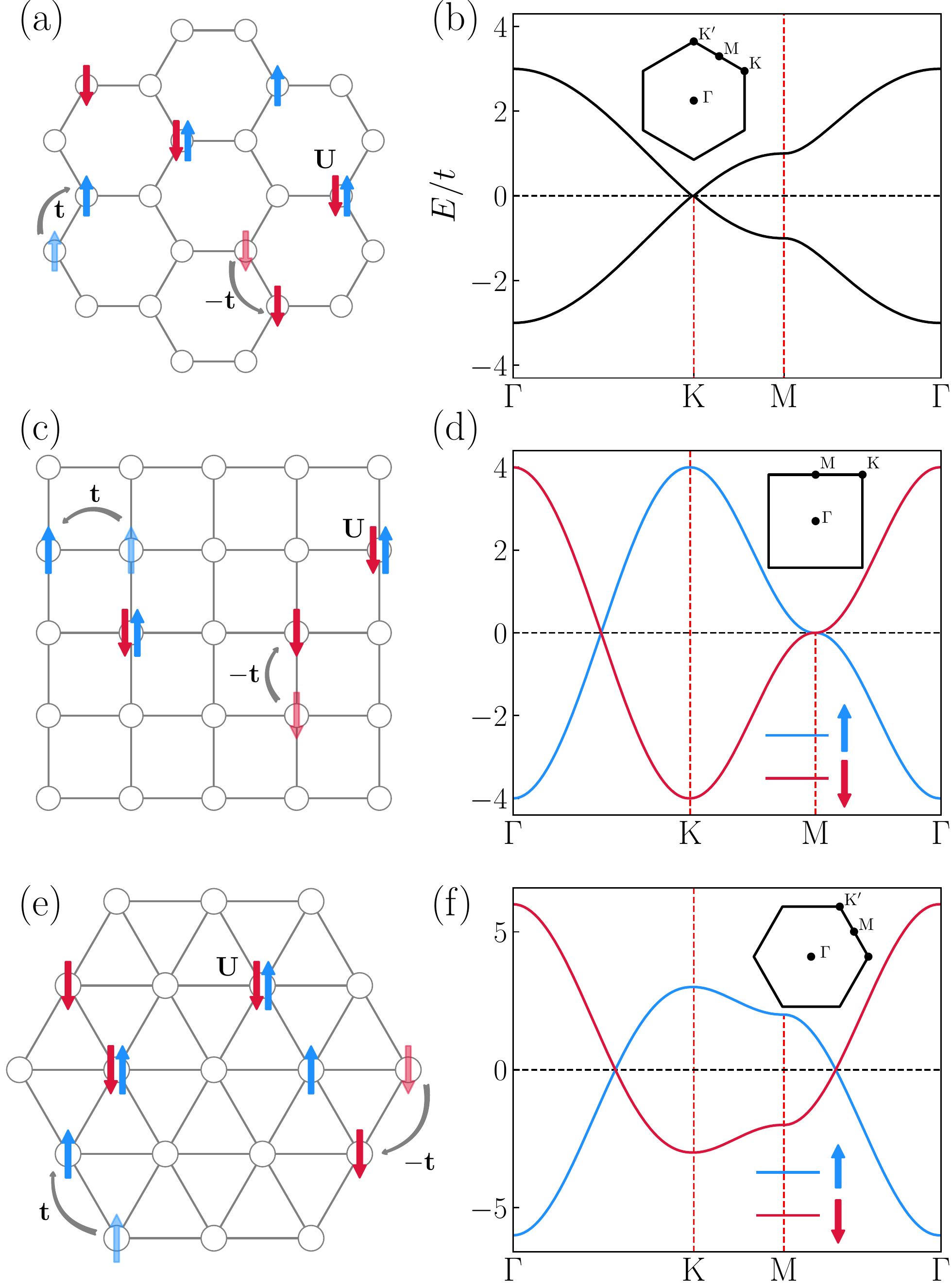} \caption{A schematic show of the Hubbard model on (a) honeycomb, (c) square, and (e) triangular lattices, where $t$($-t$) is the hopping parameter of the spin-ups (spin-downs) and $U$ is the on-site Hubbard interaction. Up and down arrows correspond to spin-up and spin-down electrons, respectively. Band structures of the imbalanced tight-binding models [the non-interacting part in Eq.(1)] on (b) honeycomb, (d) square and (f) triangular lattices. Insets in (b), (d), and (f) are the first Brillouin zones, on which the high-symmetry points are marked.}
\label{fig1}
\end{figure}

\section{Model and method}

We start from the hopping-sign imbalanced Hubbard model\cite{higher-order-kagome,otsuka2021higher,PhysRevB.105.245131},

\begin{align}\label{eq1}
H=&-t\sum_{\langle ij\rangle}\sum_{\alpha,\beta=\uparrow,\downarrow}(c^{\dagger}_{i\alpha}\sigma^z_{\alpha\beta}c_{j\beta}+\textrm{H.c.}) \\ \nonumber
&+U\sum_{i}(n_{i\uparrow}-\frac{1}{2})(n_{i\downarrow}-\frac{1}{2})-\mu \sum_{i,\alpha}n_{i\alpha}
\end{align}
where $c_{i \alpha(\beta)}^{\dagger}$ and $c_{i \alpha(\beta)}$ are the creation and annihilation operators, respectively, at site $i$ with spin $\alpha(\beta)=\uparrow, \downarrow$; $\langle ij\rangle$ denotes nearest neighbors; $\sigma^z$ is the $z$-component of Pauli matrix, and results in opposite signs in the hopping amplitudes for spin-up and -down subsystems; $n_{i \alpha}=c_{i \alpha}^{\dagger} c_{i \alpha}$ is the number operator of electrons of spin $\alpha$ on site $i$; $U$ is the on-site repulsion, and $\mu$ is the chemical potential. We set the hopping amplitude $t = 1$ as the energy scale throughout the paper.

The honeycomb lattice has a two-site unit cell [Fig.~\ref{fig1}(a)]. In momentum space, the $U=0$ Hamiltonian is spin dependent, and is given by~\cite{RevModPhys.81.109}
\begin{align}
\mathcal{H}_{0}^{\uparrow(\downarrow)}({\mathbf{k}})=\mp t\left(\begin{array}{ccc}
0 & \gamma_{\bf k} \\
\gamma^*_{\bf k} & 0
\end{array}\right),
\end{align}
where $\gamma_{\bf k}=-t \sum_j \mathrm{e}^{\mathrm{i} {\bf k} \cdot {\bf a}_j}$ with the lattice constants ${\mathbf{a}}_1=(1,0) $, and ${\mathbf{a}}_{2,3}=(-\sqrt{3},\pm 1)/2$. The spectrum has two dispersive bands $E^{\pm}_{\mathbf{k}}=\pm |\gamma_{\bf k}|$.  This noninteracting system is a semi-metal with two inequivalent Dirac points at $\boldsymbol{K}_{\pm}=\frac{2 \pi}{3}\left(1, \pm \frac{1}{\sqrt{3}}\right)$.
Three momenta $M$ at the centers of the edges of the Brillouin zone (BZ) are saddle points, resulting in the Van Hove singularities (VHSs) at the filling $\rho=3/4$ and $5/4$, respectively. Since the spectrum of itinerant electrons on honeycomb lattice is symmetric, it is not affected by the sign change in the hopping amplitudes for the spin-down subsystem, and $\mu/t=0$ still corresponds to half filling.

The square and triangular lattices are simple Bravais ones with a single crystal cell. For square lattice, the dispersion writes as $E^{\mp}_{\mathbf{k}}=\mp 2t(\cos{k_x}+\cos{k_y})$, where $-(+)$ is for the spin-up (down) band. A VHS due to the saddle point at the $M$ point of BZ is exactly located at the Fermi level, generating diverse density of states at $E/t=0$. The spectrum of triangular lattice is $E^{\mp}_{\mathbf{k}}=\mp 2t(\cos{k_x}+2\cos{\frac{k_x}{2}}\cos{\frac{\sqrt{3}k_y}{2}})$. Although the spectrum of each spin subsystem is asymmetric due to the frustration, the total one consisting of pair of opposite bands is symmetric, and half filling can be still simply achieved by setting $\mu/t=0$. Yet it is different from the bipartite situations in that the spin-up and -down densities are imbalanced at half filling with $\rho_{\uparrow}>\rho_{\downarrow}$, generating an intrinsic FM order in the $z$ direction even in the absence of interactions.

At finite interactions, Eq.\eqref{eq1} in the various geometries is solved numerically via DQMC, where one decouples the on-site interaction term through the introduction of an auxiliary Hubbard-Stratonovich field, which is integrated out stochastically~\cite{PhysRevD.24.2278,Hirsch1985,PhysRevB.40.506}. The only errors are those associated with the statistical sampling, the finite spatial lattice size, and the inverse temperature discretization. These errors are well controlled in the sense that they can be systematically reduced as needed, and further eliminated by appropriate extrapolations. While the infamous sign problem can be avoided in square and honeycomb geometries at half filling, it generally exists for the non-bipartite triangular lattice~\cite{PhysRevB.41.9301,PhysRevLett.94.170201,PhysRevB.92.045110}. Yet it is accidentally eliminated in the spin-dependent Hamiltonian Eq.(\ref{eq1}). This can be demonstrated by a simple transformation $c_{i\uparrow}\rightarrow \tilde{c}_{i\uparrow}$ and $c_{i\downarrow}\rightarrow \tilde{c}_{i\downarrow}^{\dagger}$, resulting in a normal attractive Hubbard model at half filling ($\mu/t=0$), which is free of the sign problem. When the system is doped away from half filling, the infamous sign problem usually arises due to the vanishing of the symmetries preventing the product of determinants from becoming negative. Since the sign problem becomes severe upon lowering the temperature and increasing the interaction strength, the DQMC simulations are limited to relatively high temperatures and not-so-strong interactions in the doped cases.

The magnetic order is characterized by the static structure factor, which is defined by\cite{varney2009},
\begin{align}\label{eq1a}
  S^{\alpha}({\bf k})=\sum_{\bf l}e^{i{\bf k}\cdot{\bf l}}C^{\alpha}({\bf l}),
\end{align}
where $\alpha=x,z$ denotes the spin component, and the real-space spin-spin correlation function is defined as $C^{z}({\bf l})=\langle S^z_iS^z_{i+{\bf l}}\rangle$  and $C^{x}({\bf l})=\frac{1}{2}\langle S^x_iS^x_{i+{\bf l}}+S^y_iS^y_{i+{\bf l}}\rangle$. The ferromagnetism has an order vector ${\bf k}=0$, and we let $S_{FM}^{\alpha}=S^{\alpha}({\bf k}=0)$.

To explore the intriguing superconducting properties, we explore the uniform pairing susceptibility, which is defined as~\cite{PhysRevB.91.241107,PhysRevB.97.155146,PhysRevB.97.235453},
\begin{equation}
   \chi^{\alpha}=\frac{1}{N} \int_{0}^{\beta} d \tau \sum_{i j}\left\langle\Delta_{i}^{\alpha}(\tau) \Delta_{j}^{\alpha \dagger}(0)\right\rangle,
\end{equation}
where $ \Delta_{i}^{\alpha}(\tau)= \sum_{j} f_{i j}^{\alpha} e^{\tau H} \hat{P}_{ij}^{s} e^{-\tau H}$ is the time-dependent pairing operator with form-factor $f_{i j}^{\alpha}=0, \pm 1 \text { or } \pm 2$ for the bond
connecting sites $i$ and $j$, depending on the pairing symmetry $\alpha$. Since the spins are polarized in the FM state, we consider three kinds of pairings: $\hat{P}_{ij}^{1}=c_{i \uparrow} c_{j \uparrow}, \hat{P}_{ij}^{0}=c_{i \uparrow} c_{j \downarrow}, \hat{P}_{ij}^{-1}=c_{i \downarrow} c_{j \downarrow}$, corresponding to the total $z$-spin $s_z=1,0,-1$,respectively. The effective susceptibility, $\chi_{\text {eff }}^{\alpha}\equiv\chi^{\alpha}-\chi_{0}^{\alpha}$, subtracts the uncorrelated part $\chi_{0}^{\alpha}$ from $\chi^{\alpha}$, thereby directly capturing the interaction effects, and can be further used to evaluate the pairing vertex.

In the following DQMC calculations, we use the inverse temperature discretization $\Delta\tau=0.1$, and the lowest temperature accessed is $T/t=1/20$. The lattice has $N=2\times L \times L (L \times L)$ sites for honeycomb (square and triangular) geometry with $L$ up to $20$.

\section{The magnetic transitions}

\begin{figure}[htbp]
\centering \includegraphics[width=8.5cm]{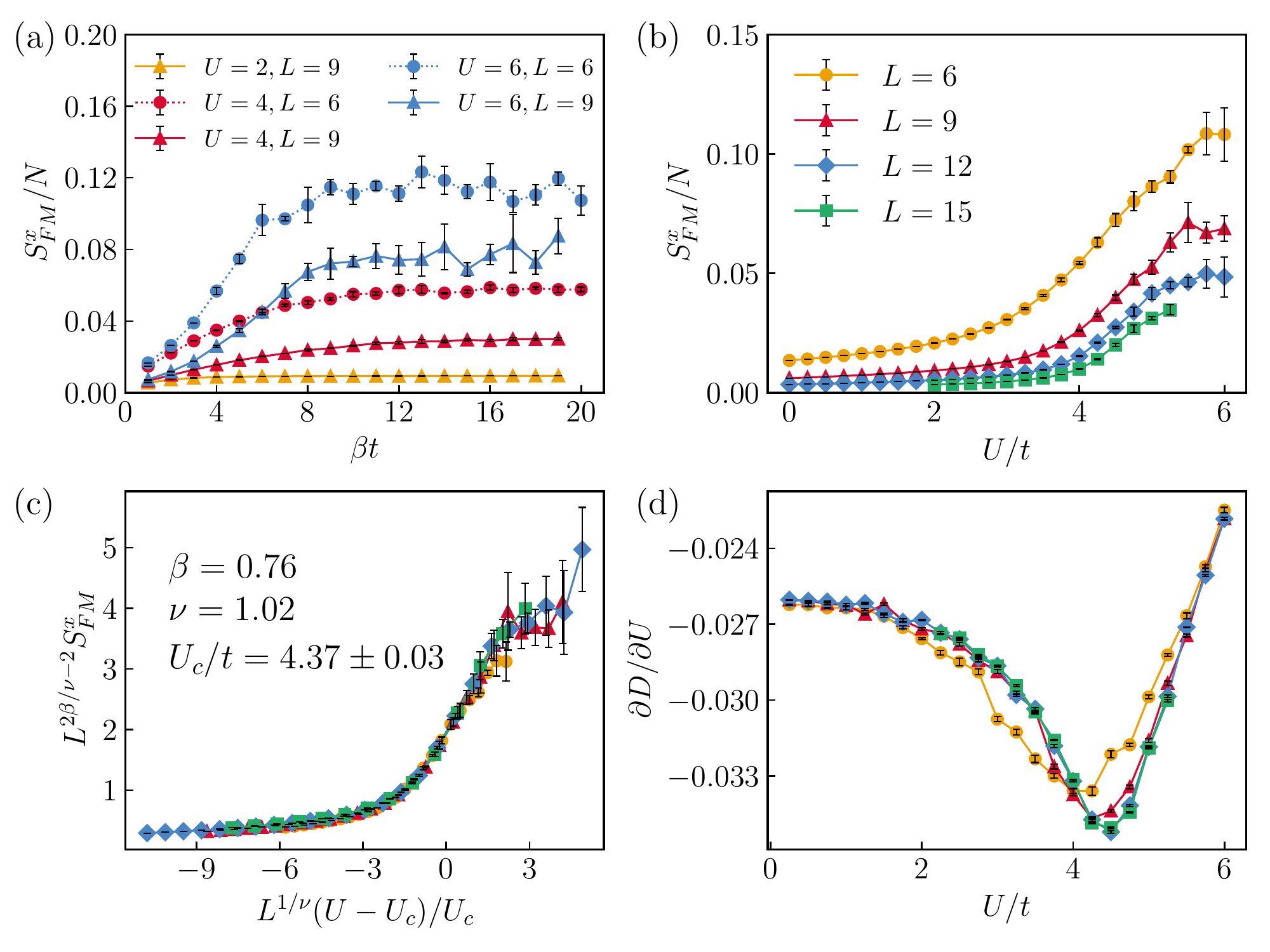} \caption{(a) The $xy$-plane FM structure factor of the imbalanced Hubbard model on a honeycomb lattice as a function of inverse temperature for various $U$ and $L$. (b) The $xy$-plane FM structure factor as a function of $U$ for various $L$ at $\beta t= 10$. (c) The best data collapse using the critical exponets of the three-dimensional Heisenberg universality class, which determines the critical interaction to be $U_c/t=4.37 \pm 0.03$. (d) The derivation of the double occupancy with respect to $U$ as a function of $U$. The $z$-direction AF structure factor is degenerate with the $xy$-plane FM one, thus is not shown here. }
\label{fig2}
\end{figure}

At half filling, the noninteracting Hamiltonian in Eq.(1) is a semimetal on honeycomb lattice. The vanishing density of states at the Fermi level suppresses the effect of the Hubbard interaction, leading to a quantum phase transition above a finite value of the on-site coupling $U$. Next we apply DQMC to unveil the magnetic transition of the Hamiltonian in Eq.(1) on honeycomb lattice quantitatively.
Figure \ref{fig2} shows the $xy$-plane FM structure factor as a function of inverse temperature for various $L$ and $U$. $S^x_{FM}$ saturates to the ground-state value at large enough $\beta$. The $z$-direction structure factor $S^{z}_{AM}$ is degenerate with $S^x_{FM}$, thus is not demonstrated here. Although a larger value of $\beta$ is required as $L$ increases, $\beta t=10$ is low enough to represent the property of the ground state for the accessed lattice sizes here, and is adopted in the following DQMC simulations.

To see the evolutions of the structure factors with $U$, we show $S^{x}_{FM}$ versus the interaction strength for various $L$ at $\beta t=10$ in Fig.\ref{fig2} (b). As $U$ increases, $S^{x}_{FM}$ increases monotonically.
Their values remain negligibly small for weak interacting strength, and becomes finite only for large enough $U$. This behavior indicates the magnetic order occurs above a finte interaction, which is consistent with the previous qualitative analysis.
We perform a finite-size scaling analysis of $S_{FM}^{x}$ at different lattice sizes based on the following commonly used scaling formula\cite{assaad2013},
\begin{align}
S_{FM}^{x}=L^{2-2\beta/\nu} F[L^{1/\nu}(U-U_c)],
\end{align}
where $\beta$ is the order parameter exponent, and $\nu$ is the correlation length exponent.
The magnetic transition is expected to belong to the three-dimensional Heisenberg universality class. Various methods have been applied to investigate the interaction-driven phase transition in interacting Dirac fermions in $d=2+1$ based on the
honeycomb and $\pi$-flux lattice models, and the effective continuous Gross-Neveu models with the total number
$N=8$ of fermion components. Although the estimated critical exponents are consistent among the existing studies, they still differ from one another slightly. Here we use the values $\beta=0.76$ and $\nu=1.02$, which is obtained by a recent large-scale DQMC simulations\cite{PhysRevX.6.011029}, and take $U_c$ as a fitting parameter.
As shown in Fig.\ref{fig2}(d), the best data collapse is obtained at $U_c/t=4.37 \pm 0.03$ for the above fixed values of $\beta, \nu$. Thus along with determining the critical interaction, the perfect data collapse further confirms that the phase transition belongs to the three-dimensional Heisenberg universality class\cite{zinn2021quantum}.

Our simulations also find the double occupancy $D=\langle n_{i\uparrow}n_{i\downarrow}\rangle$ continuously
decreases with $U$, which is expected since the on-site repulsion $U$ suppresses the double occupancy. Remarkably, accompanying the magnetic transition, there appears a peak in the absolute value of the derivation of the double occupancy [see Fig.\ref{fig2}(d)], where $D$ is decreasing most rapidly. It indicates the phase transition may also manifest itself as an anomaly in the double occupancy\cite{PhysRevB.96.205130,PhysRevLett.103.036401,mondaini2022quantum}.

The coexistence of the FM and AF magnetic orders can be easily understood in the large-$U$ limit, when the double occupancy is completely eliminated, and the Hubbard model in Eq.(1) maps onto the following Heisenberg model\cite{charles1976},
\begin{align}
  {\cal H}=-J\sum_{\langle ij\rangle} (S^x_i S^x_j+S^y_i S^y_i)+J\sum_{\langle ij\rangle} S^z_iS^z_j,
\end{align}
where the exchange coupling is $J=\frac{4t^2}{U}$. Due to the different sighs of the exchange couplings in the three directions, the system exhibits FM (AF) order when the spontaneous symmetry breaking occurs in the $xy$ plane ($z$ direction). In addition, the spin-$\frac{1}{2}$ operators can be mapped to hardcore-boson ones via $S_i^+=a_i^{\dagger}, S_i^-=a_i, S_i^z=n_i-1/2$ ($n_i=a_i^{\dagger}a_i$), where $a^{\dagger}_i$ and $a_i$ are the hardcore-boson creation and annihilation operators, respectively\cite{10.1143/PTP.16.569}. The resulting hardcore Bose-Hubbard model on honeycomb lattice writes as,
\begin{align}
H_{BH}=\sum_{\langle ij\rangle}\left[-t(a_i^{\dagger} a_j+\textrm{H.c.}\right)+V n_i n_j]-\mu \sum_i n_i,
\end{align}
with $t$ the hopping amplitude, $V$ the nearest-neighbor interaction and $\mu$ the chemical potentia. The above Hamiltonian
has been widely investigated in the literature\cite{Isakov2006,Isakov2006b,Isakov2007}. For the parameter values in Eq.(6), we have $t=J/2$, $V=2t, \mu=3t$. This set of parameters corresponds to the Heisenberg point of the phase diagram, which is located at the tip of the lobe phase boundary between the superfulid and charge-density-wave states.

\begin{figure}[htbp]
\centering \includegraphics[width=8.5cm]{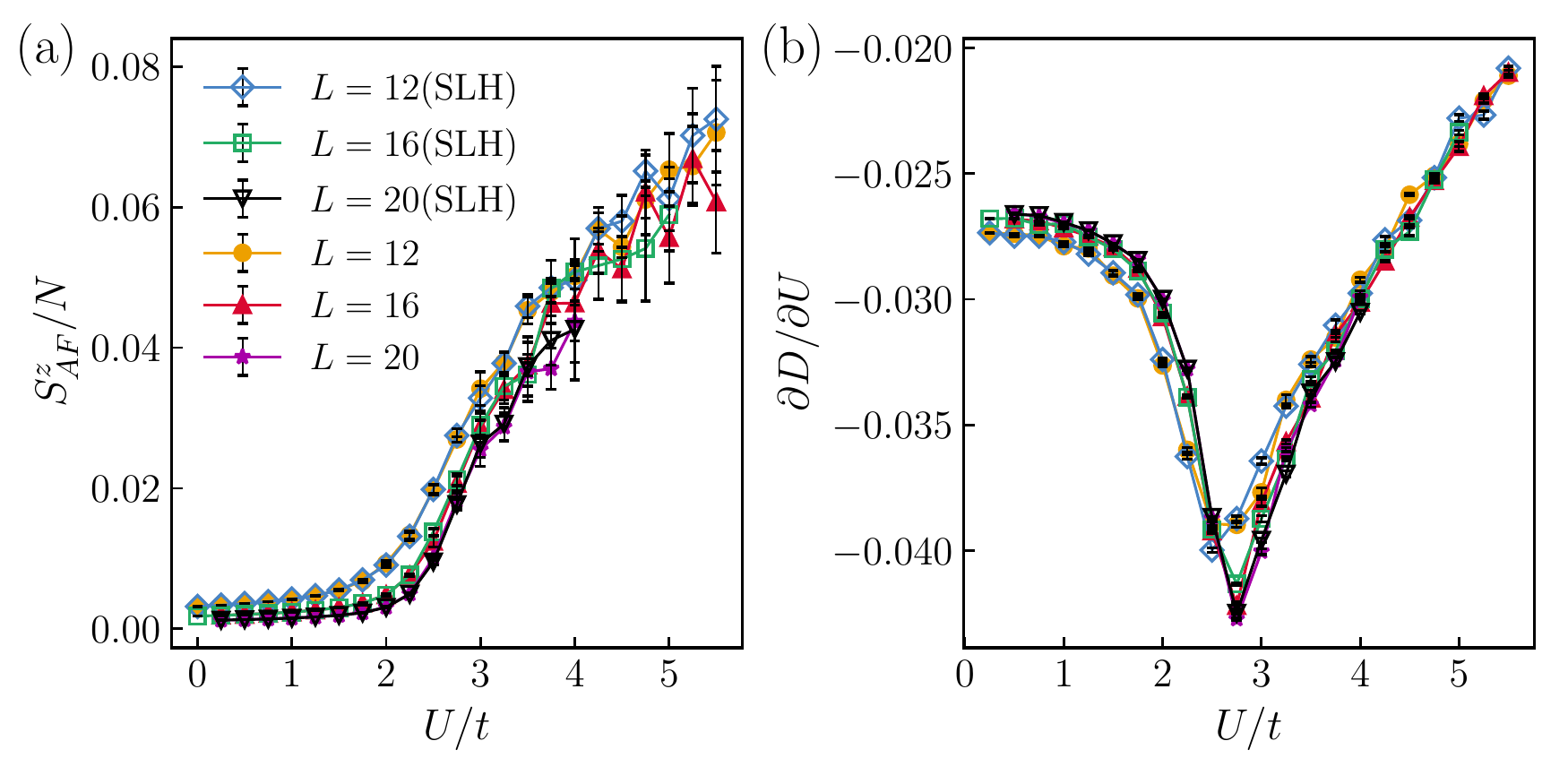} \caption{(a) The $z$-direction AF structure factor as a function of $U$ for various $L$ at $\beta t= 10$ in the imbalanced square-lattice Hubbard model. (b) The derivation of the double occupancy with respect to $U$ as a function of $U$.  For comparison, the corresponding quantities in the normal square-lattice Hubbard model(SLH) are also demonstrated.}
\label{fig3}
\end{figure}

In the normal square-lattice Hubbard model, it has been well established that the perfect-nesting instability towards antiferromagnetism can occur at any infinitesimal $U$\cite{Hirsch1985,qin2022hubbard}. However in actual numerical finite-size scalings, it is very hard to extrapolate to $U_c=0$, which is also the situation we encounter in dealing with the magnetic transition in the imbalanced square-lattice Hubbard model. Therefore, we compare the spin structure factor and the derivation of the double occupancy in the normal and imbalanced Hubbard Hamiltonians. As shown in Fig.\ref{fig3}, both physical quantities are  equal to the statistical error. So although a finite $U_c$ is identified here on several lattice sizes for the imbalanced Hamiltonian, it should be due to the finite-size effect, and $U_c=0$ should be expected in the thermodynamic limit.

\begin{figure}[htbp]
\centering \includegraphics[width=8.5cm]{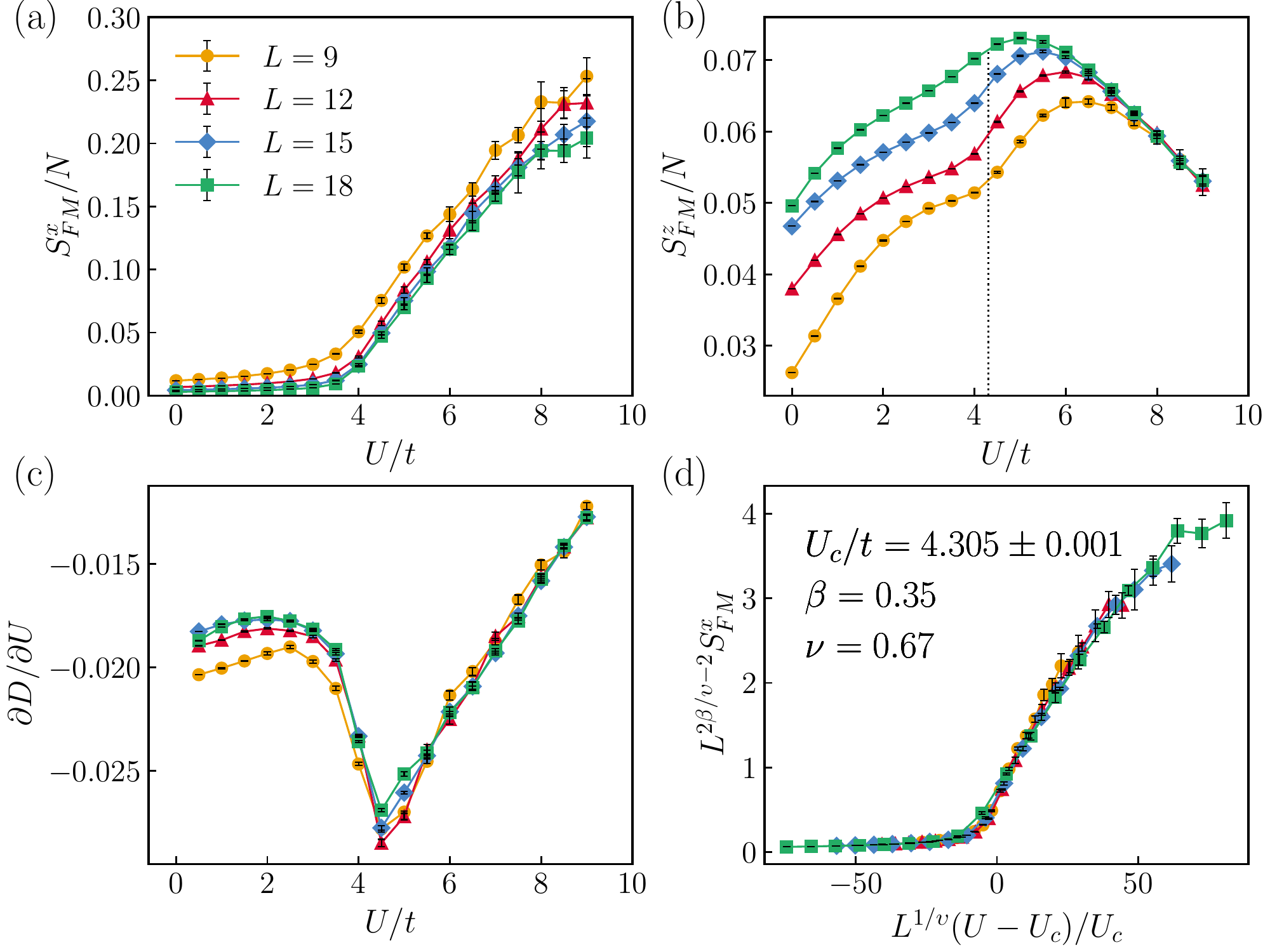} \caption{The $xy$-plane (a) and $z$-direction (b) FM structure factors as a function of $U$ for various $L$ at $\beta t = 10$ in the imbalanced triangle-lattice Hubbard model. (c) The derivation of the corresponding double occupancy with respect to $U$. (d) The best data collapse using the critical exponets
of the three-dimensional $XY$ universality class, which determines the critical interaction to be $U_c/t = 4.305 \pm 0.001$. The dashed vertical line in (b) marks the transition point.}
\label{fig4}
\end{figure}

Unlike the square and honeycomb lattices, the triangular one is non-bipartite, and the band structure is asymmetric. The induced imbalance of the numbers of spin-up and -down electrons at half filling results in an intrinsic $z$-direction FM order even at $U/t=0$. As shown in Fig.\ref{fig4}(b), $S^{z}_{FM}$ evolutes non-monotonically with $U$, and exhibits a peak at a moderate value of $U$, after which $S^{z}_{FM}$ decreases continuously. It tends to vanish in the large-$U$ limit, which can be well understood in terms of the mapped Bose-Hubbard model: at the corresponding parameters, the system is a superfluid, which in fact is a $xy$-plane FM order in the spin language. The curves of $S^{x}_{FM}$ as a function of $U$ are similar to those on the honeycomb lattice shown in Fig.\ref{fig2}(b), and the FM order in the $xy$ plane occurs at a critical interaction. An anomaly is also found in the derivation of the double occupancy at the transition point [see Fig.\ref{fig4}(c)]. This $xy$-plane FM transition belongs to the three-dimensional $XY$ universality class. With the known critical exponents $\beta=0.35$ and $\nu=0.67$, the critical interaction is estimated to be $U_c=4.305\pm 0.001$ by the best collapse of the curves for different lattice sizes[Fig.\ref{fig4}(d)].

\section{The $p$-wave triplet superconductivity in the imbalanced Hubbard model}

In the systems with electron-electron interactions, the spin fluctuation plays an important role in mediating the superconducting pairing. While the AF fluctuations favor unconventional spin-singlet pairing, the FM ones may lead to novel spin-triplet superconductivity. Our system has long-range $xy$-plane FM order at half filling, thus the spin-triplet pairing is highly expected to occur in a doped system. To reveal the dominating superconducting instability, we calculate the effective susceptibility of all possible pairing channels as a function of temperature for various values of $U$ at $\rho=0.95$. It should be noted that in the doped region, the sign problem occurs, and the DQMC simulations are limited to relatively high temperatures. So only the high-temperature trends of the pairing susceptibility can be obtained, which is still informative in determining the dominative pairing instability.

\begin{figure}[htbp]
\centering \includegraphics[width=6.5cm]{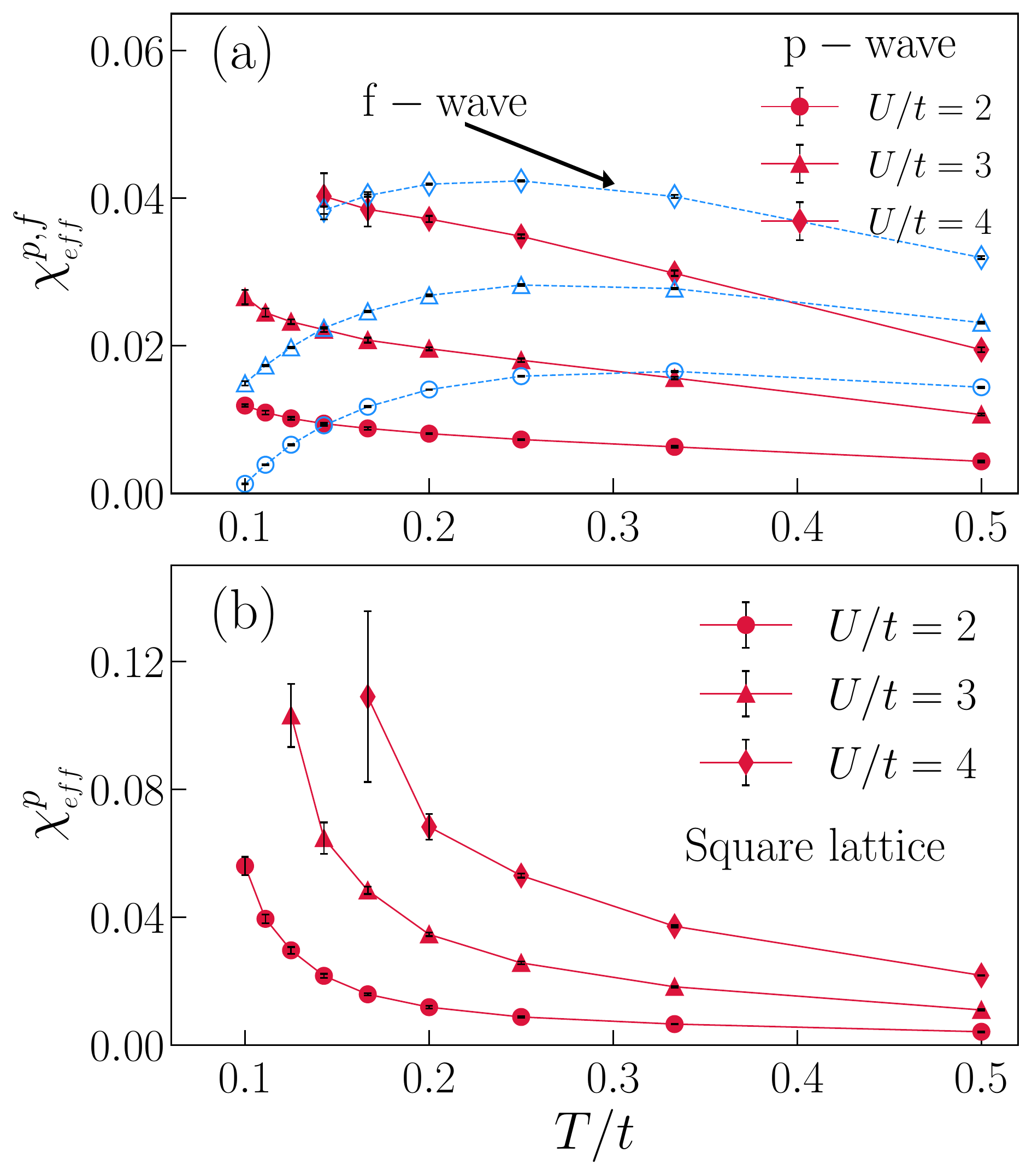} \caption{The effective susceptibility of the triplet pairing
channels as a function of temperature for several values of $U$ in (a) a honeycomb lattice and (b) a square lattice. Here the system is at $5\%$ hole doping, corresponding to the average density $\rho=0.95$.}
\label{fig5}
\end{figure}

\begin{figure}[htbp]
\centering \includegraphics[width=8.5cm]{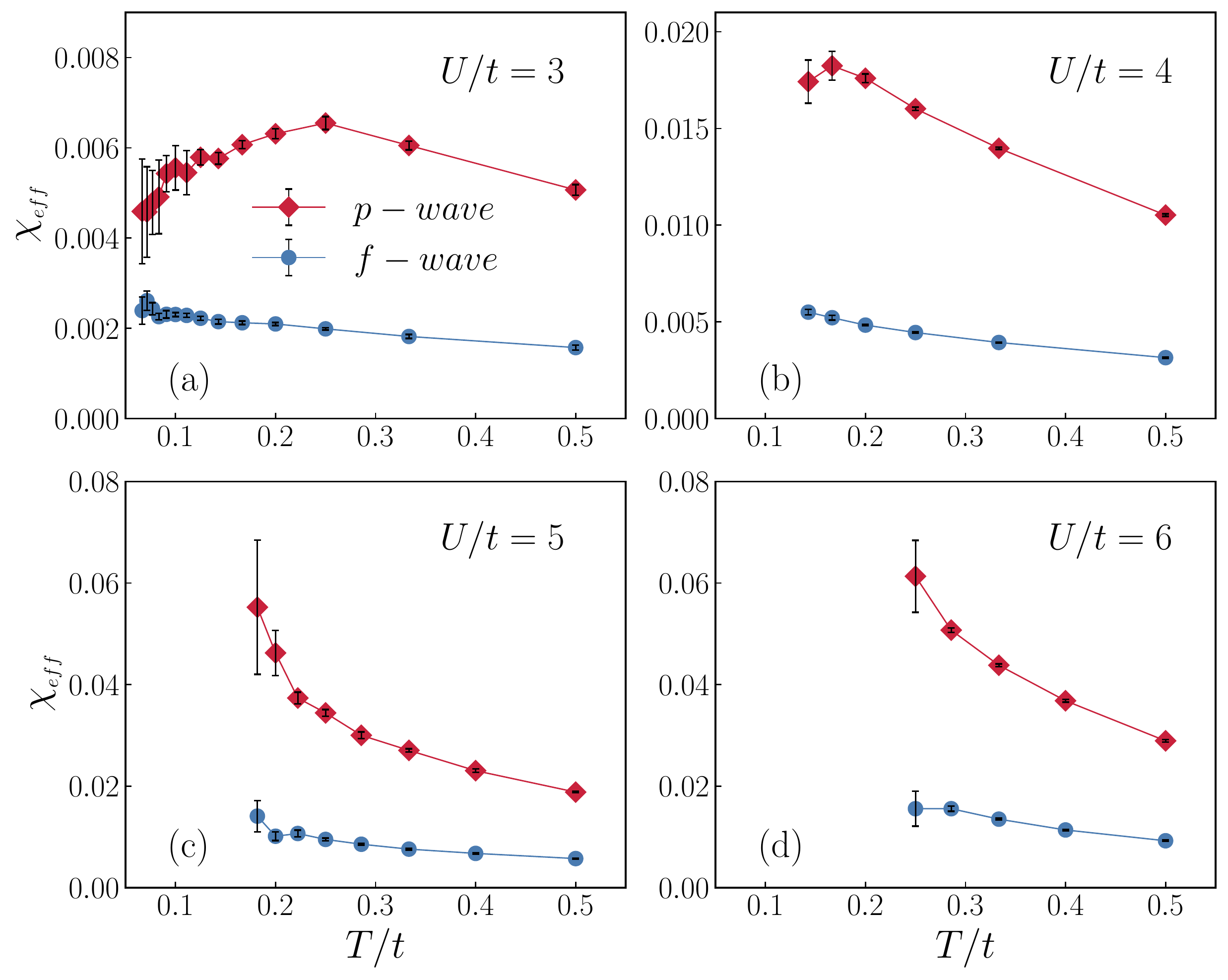} \caption{The effective susceptibility of the triplet pairing
channels as a function of temperature in the imbalanced triangle-lattice Hubbard model for (a) $U/t=3$, (b) $U/t=4$, (c) $U/t=5$, and (d) $U/t=6$. Here the average density is the same with that in Fig.\ref{fig5}, i.e., $\rho=0.95$.}
\label{fig6}
\end{figure}

We first consider the honeycomb geometry.  Its crystal symmetry group is $C_{6v}$, the irreducible representation of which can classify the possible pairing states. The allowed pairing symmetries include: singlet $s^*$-wave,$d_{x^2-y^2}$-wave, $d_{x y}$-wave; triplet $p_x$-wave, $p_y$-wave, $f$-wave\cite{Black-Schaffer_2014,nandkishore2012chiral,PhysRevB.90.245114,PhysRevB.106.134513,Zhu_2019}. Among them, the $d_{x^2-y^2}$ and $d_{x y}$ ($p_x$ and $p_y$) channels are degenerate since they belong to the same two-dimensional representation. As shown in Fig.5(a), while $\chi_{eff}^{p}$ and $\chi_{eff}^{f}$ exhibit clear enhancements over their uncorrelated values (implying the corresponding pairing interactions are attractive), the values of $s^{*}$-, $d$-wave pairings are negative, and decrease with increasing the interaction (see Appendix A). This indicates the effective paring interactions therein are repulsive, and the singlet channels are continuously suppressed by the interactions. The curve of $\chi_{eff}^{f}$ exhibits a peak above the superconducting critical temperature, below which $\chi_{eff}^{f}$ begins to decrease. In contrast, $\chi_{eff}^{p}$ grows monotonically with lowering the temperature, and has a trend to diverge at a lower temperature.
Hence, the possible superconducting instability of the ground state should have a $p$-wave symmetry.
In addition, we find the value of $\chi_{eff}^{p}$ increases with the interaction, thus the superconductivity is gradually enhanced by the interaction.
We also calculate the effective susceptibility of the $s_z = \pm 1$ pairings. It is found the values for the $p$-wave pairing are increasingly negative with decreasing the temperature, suggesting these finite-$s_z$ paring channels are disfavoured.

For the imbalanced square-lattice Hubbard model, the singlet $s^*$-wave, $d_{x^2-y^2}$-wave, and triplet $p_x$-wave, $p_y$-wave nearest-neighbor pairings are considered\cite{PhysRevB.97.155146,PhysRevB.91.241107}. The corresponding pair functions in the momentum space are,
\begin{align}
f_{\mathbf{k}}\left(s^*\right)=\cos k_x+\cos k_y, \quad f_{\mathbf{k}}\left(p_x\right) & =\sin k_x, \\
f_{\mathbf{k}}\left(d_{x^2-y^2}\right)=\cos k_x-\cos k_y, \quad f_{\mathbf{k}}\left(p_y\right) & =\sin k_y.
\end{align}
Among the different symmetries, we find only $\chi_{eff}^{p}$ is positive, and tends to diverge at the superconducting critical temperature. Thus by changing the hopping sign of the spin-down subsystem, we realize a $p$-wave triplet superconducting ground state in the square-lattice Hubbard model.

\begin{figure}[htbp]
\centering \includegraphics[width=8.5cm]{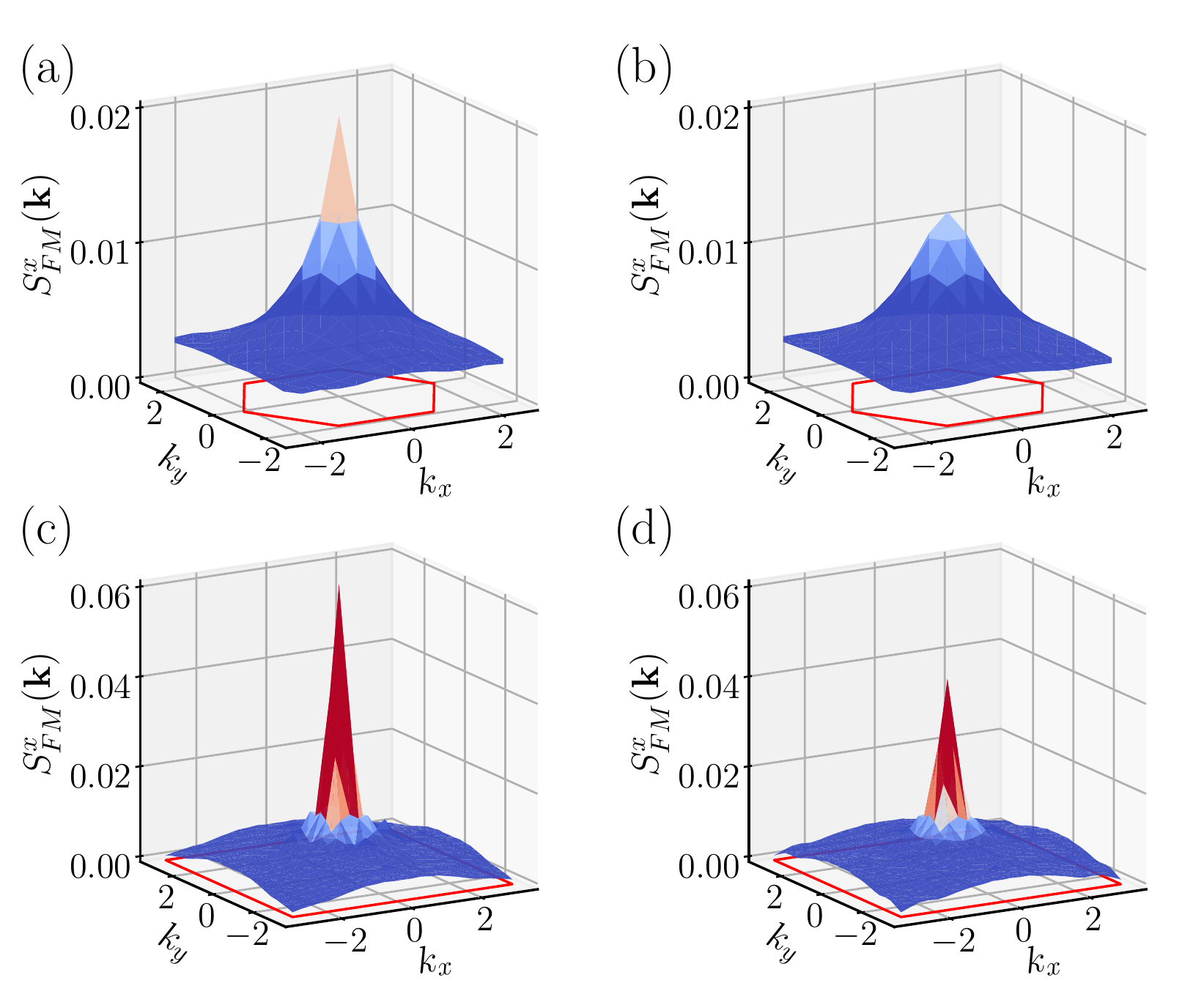} \caption{The transverse FM structure factors at $U/t=4$ on honeycomb lattice: (a) half filling ($\rho=1$); (b) $5\%$ hole doping ($\rho=0.95$). (c) and (d) present the results of square lattice, corresponding those in (a) and (b), respectively. The longitudinal AF structure factors are the same as the transverse ones, thus are not shown here. The inverse temperature is $\beta t=6$.}
\label{fig7}
\end{figure}

\begin{figure}[htbp]
\centering \includegraphics[width=8.5cm]{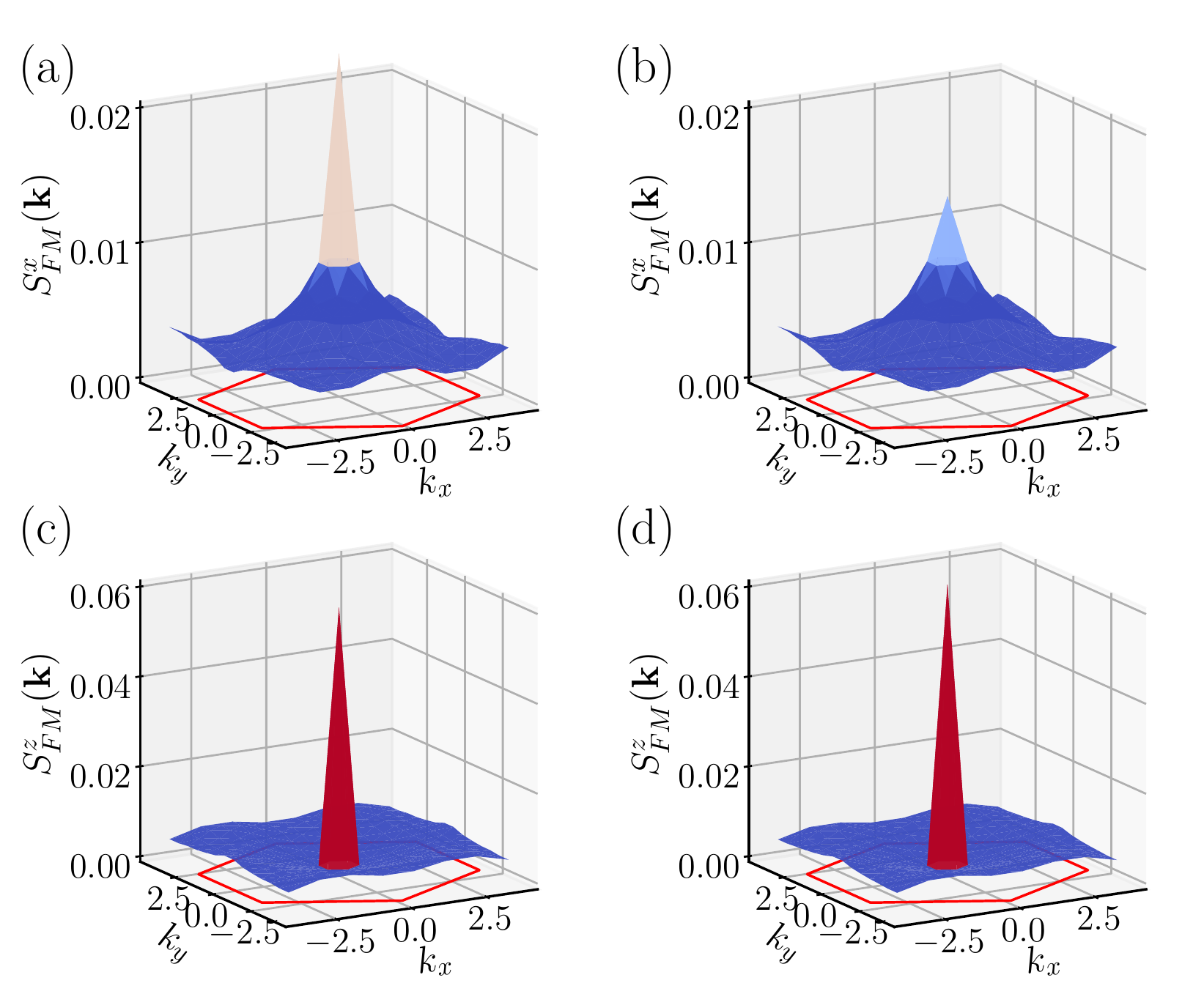} \caption{The transverse FM structure factors at $U/t=4$ on triangular lattice: (a) half filling ($\rho=1$); (b) $5\%$ hole doping ($\rho=0.95$). (c) and (d) plot longitudinal ones with the fillings in (a) and (b), respectively.}
\label{fig8}
\end{figure}

Since the triangular lattice also belongs to the space group $C_{6v}$, the possible pairing channels are the same with those of the honeycomb lattice\cite{PhysRevB.97.235453,PhysRevB.88.041103}. We find that only the effective susceptibility of the triplet pairings are positive. While $\chi_{eff}^{p}$ begins to drop at low temperatures for weak interactions, it has a trend to be divergent at large interactions. This suggests an instability to the $p$-wave superconductivity in the triangle-lattice Hubbard model at large $U$. In the weak-coupling region, $\chi_{eff}^{f}$ is much smaller than $\chi_{eff}^{p}$, and increases slowly with lowering $T$. From this high-temperature behavior, it is still insufficient to determine whether the $f$-wave channel will dominate in the ground state.

To reveal the microscopic origin of the superconducting pairing interaction, we calculate the spin correlations, and compare the values at half filling ($\rho=1$) with those of the hole-doped case ($\rho=0.95$).  For the square and honeycomb lattices, we find that the spin correlations of all three components decrease quickly with the distance, and become short-ranged after the holes are doped. Figure 5 plots the static spin structure factor of the $xy$-plane FM order. The sharp FM peak at ${\bf k}=(0,0)$ at half filling is greatly suppressed by the doping. Due to the degeneracy, the evolution of $S^z_{FM}({\bf k})$ describing the $z$-direction AF order is exactly the same, thus is not plotted here. The short-range spin correlations will generate strong spin fluctuations, which can mediate the superconducting pairings. Since the transverse FM spin fluctuations are contributed by two spin components, they will dominate the pairing interaction, resulting in a triplet superconductivity. This is consistent with the enhancement of the triplet $s_z=0$ superconductivity revealed by the above trend of the effective susceptibility, thus suggesting the triplet superconductivity may be mediated by the $xy$-plane FM fluctuations. This also naturally explains the absence of finte-$s_z$ ($s_z=\pm 1$) triplet pairings.
Although the $z$-direction AF spin fluctuations also exist in the hole-doped system, they are weaker than the FM ones, and the singlet channels will not be generated.

The situation in the triangular lattice is different, where both the transverse and longitudinal magnetic properties are ferromagnetic, and they are no longer degenerate. It is found that while $S^x_{FM}({\bf k}=0)$ is greatly suppressed in the doped system, $S^z_{FM}({\bf k}=0)$ is almost unchanged by the doped holes. Thus the triplet superconducting pairing is mediated by the pronounced transverse spin fluctuations induced by the breakdown of the long-range FM order in the $xy$ plane.

\section{Conclusions}

We have applied the DQMC simulations to study the magnetic transition and the superconducting pairing symmetry in the imbalanced Hubbard model on honeycomb, square, and triangular lattices. For the bipartite geometries, the magnetic property is FM and AF in the $xy$ plane and the $z$ direction, respectively. While the magnetism can occur at any finite $U$ on square lattice, there is a critical interaction for honeycomb lattice, which is estimated to be $U_c=4.37\pm 0.03$ by finite-size scaling. Unlike the above two bipartite lattices, the triangular lattice is frustrated, and the asymmetry band structure leads to the imbalance of the spin-up and -down electron densities, resulting in a $z$-direction FM order inherently at half filling. We find that the $xy$-plane ferromagnetism develops above a critical interaction, which is estimated to be $U_c/t=4.305\pm 0.001$. Both the magnetic transitions on honeycomb and triangular lattices are continuous, and are verified to belong to the three-dimensional Heisenberg and $XY$ universality classes, respectively. We then investigate the pairing symmetry of the superconducting instability in the doped system. From the low-temperature trend of the effective pairing susceptibility, we unveil that a triplet $p$-wave pairing will be dominant in the possible superconducting ground state. Our study provides a universal approach to obtain $p$-wave triplet superconductivity, which will not only deepen the understanding of the microscopical mechanism of the triplet pairing, but also be helpful in guiding the exploration of the triplet superconducting materials.

Recently, a new magnetic phase called altermagnetism is discovered in a number of magnetic materials\cite{PhysRevX.12.031042,PhysRevB.99.184432,PhysRevX.12.040501}, such as: $\rm{Ru}\rm{O}_2$, $\rm{K}\rm{Ru}_{4}\rm{O}_8$, $\rm{Mn}_5\rm{Si}_3$, et.al.. The electron quasiparticle therein is described by a hopping-sign imbalanced Hamiltonian similar to the one studied here\cite{PhysRevX.12.040501}. Hence, it will be promising that our theoretical results are explored in these new altermagnetic materials.

\section*{Acknowledgments}
The authors thank Song-Bo Zhang for helpful discussions. W.H. and H.G. acknowledge support from the National Natural Science Foundation of China (NSFC) grant Nos.~11774019 and 12074022, the NSAF grant in NSFC with grant No. U1930402.
S.F. is supported by the National Key Research and
Development Program of China under Grant No. 2021YFA1401803,
and NSFC under Grant Nos. 11974051 and 12274036.

\appendix

\renewcommand{\thefigure}{A\arabic{figure}}

\setcounter{figure}{0}

\section{The effective susceptibilities of the spin-singlet pairings}

\begin{figure}[htbp]
\centering \includegraphics[width=7.5cm]{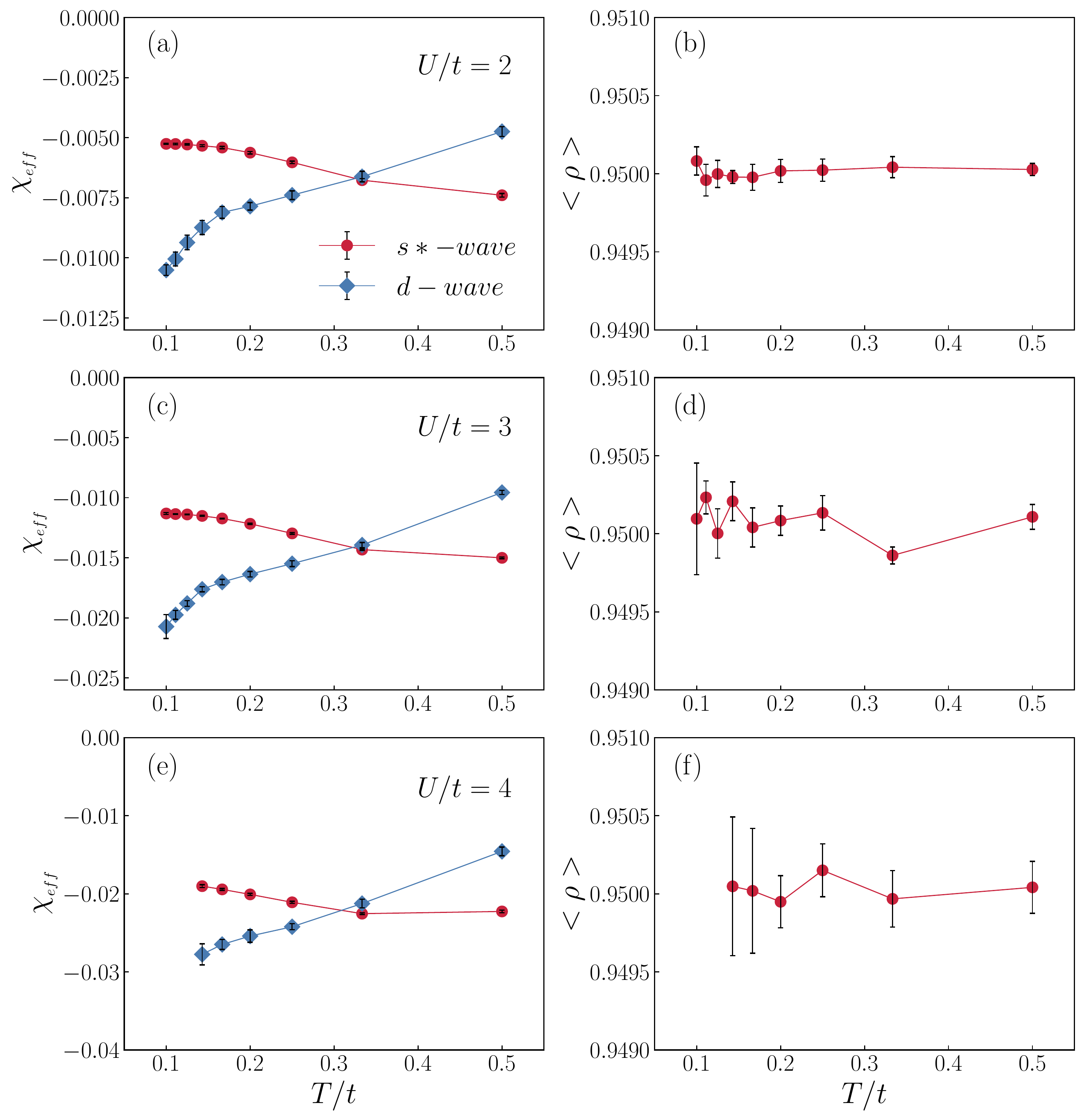} \caption{The effective pairing susceptibility of $s^{*}$- and $d$-wave
channels as a function of temperature on honeycomb lattice for: (a) $U/t=2$; (c) $U/t=3$; (e) $U/t=4$. Here the system is at $5\%$ hole doping. (b), (d), and (f) show the average density at the manually determined chemical potential targeting the fixed density $\rho=0.95$ at the interaction strengths in (a), (c), and (e), respectively.}
\label{afig1}
\end{figure}

\begin{figure}[htbp]
\centering \includegraphics[width=7.5cm]{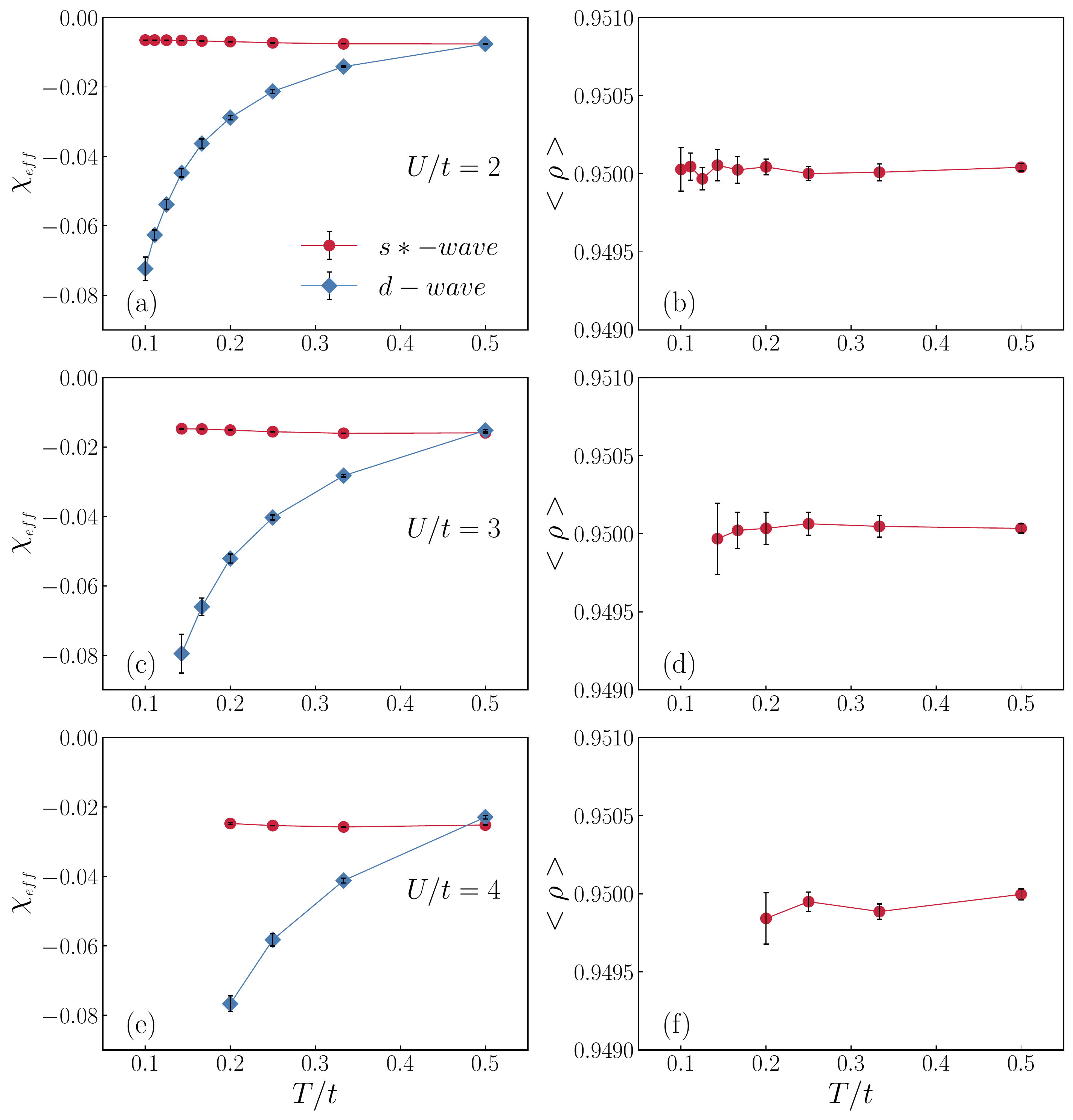} \caption{The effective pairing susceptibility of $s^{*}$- and $d$-wave
channels as a function of temperature on square lattice for: (a) $U/t=2$; (c) $U/t=3$; (e) $U/t=4$. Here the system is at $5\%$ hole doping. (b), (d), and (f) show the average density at the manually determined chemical potential targeting the fixed density $\rho=0.95$ at the interaction strengths in (a), (c), and (e), respectively.}
\label{afig2}
\end{figure}

\begin{figure}[htbp]
\centering \includegraphics[width=7.5cm]{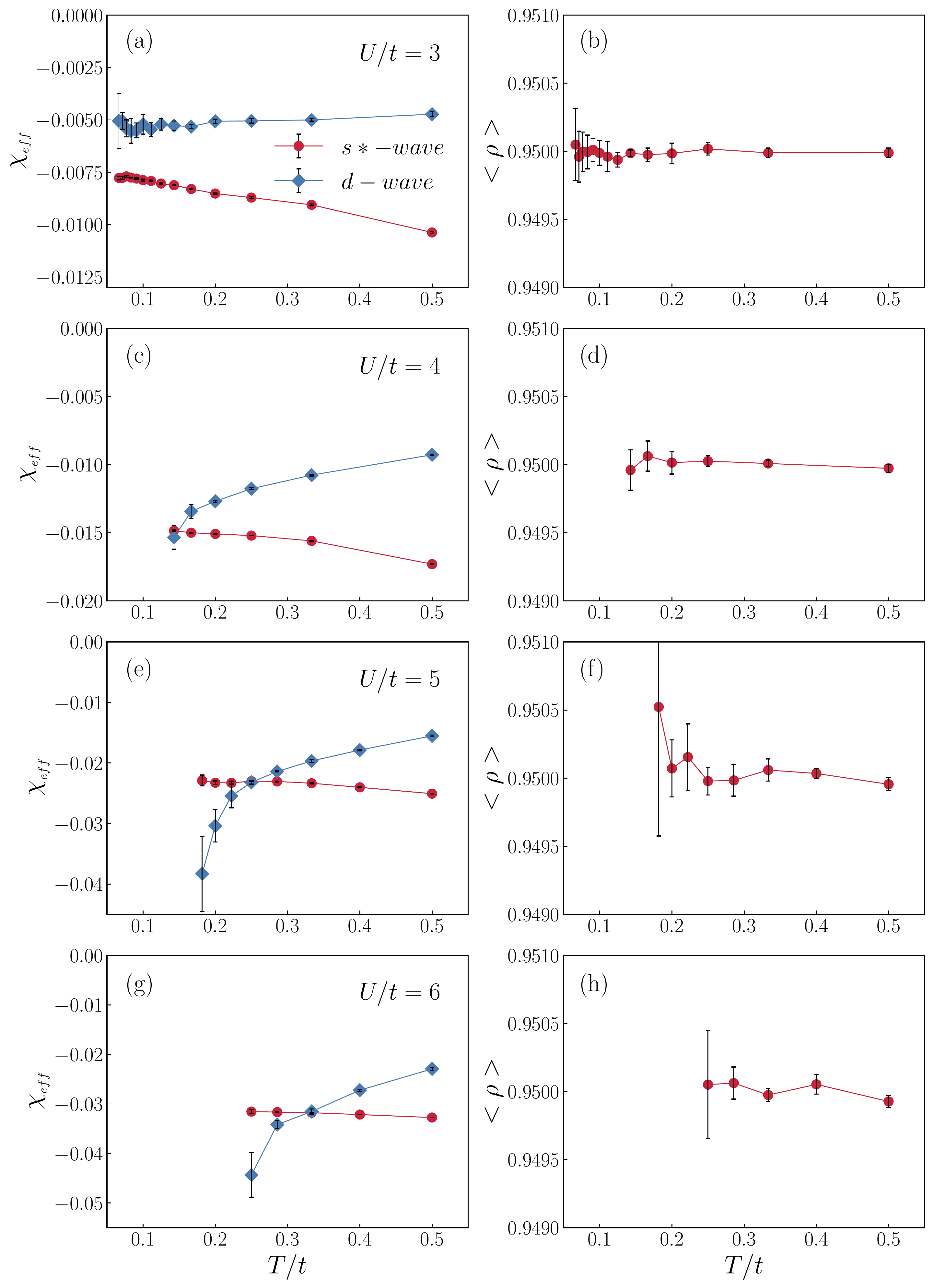} \caption{The effective pairing susceptibility of $s^{*}$- and $d$-wave
channels as a function of temperature on triangular lattice for: (a) $U/t=3$; (c) $U/t=4$; (e) $U/t=5$; (g) $U/t=6$. Here the system is at $5\%$ hole doping. (b), (d), (f), and (h) show the average density at the manually determined chemical potential targeting the fixed density $\rho=0.95$ at the interaction strengths in (a), (c), (e) and (g), respectively.}
\label{afig3}
\end{figure}

In the main text we have demonstrated the pairing susceptibilities of the dominating spin-triplet pairings in the hole-doped case. Here we provide further justification, by investigating the spin-singlet channels, in complement to Fig.5 and Fig.6 in the main text. Figure \ref{afig1}-\ref{afig3} show the effective susceptibilities for $s^*$- and $d$-wave singlet pairings. For all the three considered lattices, $\chi_{eff}^{s^*,d}$ is negative over the temperature range simulated by DQMC, which suggest the above two singlet pairing symmetries are suppressed by the on-site Hubbard interaction.

\bibliography{ddirac}


\end{document}